


\font\titlefont = cmr10 scaled\magstep 4
\font\sectionfont = cmr10
\font\littlefont = cmr5
\font\eightrm = cmr8

\def\ss{\scriptstyle}
\def\sss{\scriptscriptstyle}

\newcount\tcflag
\tcflag = 0  

\ifnum\tcflag = 0 \magnification = 1200 \fi  

\global\baselineskip = 1.2\baselineskip
\global\parskip = 4pt plus 0.3pt
\global\abovedisplayskip = 18pt plus3pt minus9pt
\global\belowdisplayskip = 18pt plus3pt minus9pt
\global\abovedisplayshortskip = 6pt plus3pt
\global\belowdisplayshortskip = 6pt plus3pt

\def\barsoff{\overfullrule=0pt}


\def\endignore{}
\def\ignore #1\endignore{}

\newcount\dflag
\dflag = 0


\def\monthname{\ifcase\month
\or January \or February \or March \or April \or May \or June%
\or July \or August \or September \or October \or November %
\or December
\fi}

\newcount\dummy
\newcount\minute  
\newcount\hour
\newcount\localtime
\newcount\localday
\localtime = \time
\localday = \day

\def\advanceclock#1#2{ 
\dummy = #1
\multiply\dummy by 60
\advance\dummy by #2
\advance\localtime by \dummy
\ifnum\localtime > 1440 
\advance\localtime by -1440
\advance\localday by 1
\fi}

\def\settime{{\dummy = \localtime%
\divide\dummy by 60%
\hour = \dummy
\minute = \localtime%
\multiply\dummy by 60%
\advance\minute by -\dummy
\ifnum\minute < 10
\xdef\spacer{0} 
\else \xdef\spacer{}
\fi %
\ifnum\hour < 12
\xdef\ampm{a.m.} 
\else
\xdef\ampm{p.m.} 
\advance\hour by -12 %
\fi %
\ifnum\hour = 0 \hour = 12 \fi
\xdef\timestring{\number\hour : \spacer \number\minute%
\thinspace \ampm}}}



\def\endtitle{}
\def\title#1\endtitle{\vskip.5in\titlefont
\global\baselineskip = 2\baselineskip
#1\vskip.4in
\baselineskip = 0.5\baselineskip\rm}

\def\endauthors{}
\def\authors#1\endauthors{#1}

\def\endabstract{}
\def\abstract#1\endabstract{\vskip .3in%
\centerline{\sectionfont\bf Abstract}%
\vskip .1in
\noindent#1}

\def\nopageonenumber{\footline={\ifnum\pageno<2\hfil\else
\hss\tenrm\folio\hss\fi}}  

\newcount\nsection
\newcount\nsubsection

\def\section#1{\global\advance\nsection by 1
\nsubsection=0
\bigskip\noindent\centerline{\sectionfont \bf \number\nsection.\ #1}
\bigskip\rm\nobreak}

\def\subsection#1{\global\advance\nsubsection by 1
\bigskip\noindent\sectionfont \sl \number\nsection.\number\nsubsection)\
#1\bigskip\rm\nobreak}

\def\topic#1{{\medskip\noindent $\bullet$ \it #1:}}
\def\endtopic{\medskip}

\def\appendix#1#2{\bigskip\noindent%
\centerline{\sectionfont \bf Appendix #1.\ #2}
\bigskip\rm\nobreak}


\newcount\nref
\global\nref = 1

\def\therefs{}


\def\ref#1#2{\xdef #1{[\number\nref]}
\ifnum\nref = 1\global\xdef\therefs{\item{[\number\nref]} #2\ }
\else
\global\xdef\oldrefs{\therefs}
\global\xdef\therefs{\oldrefs\vskip.1in\item{[\number\nref]} #2\ }%
\fi%
\global\advance\nref by 1
}

\def\listrefs{\vfill\eject\section{References}\therefs}


\newcount\nfoot
\global\nfoot = 1

\def\foot#1#2{\xdef #1{(\number\nfoot)}
\footnote{${}^{\number\nfoot}$}{\eightrm #2}
\global\advance\nfoot by 1
}


\newcount\nfig
\global\nfig = 1
\def\thefigs{} 

\def\figure#1#2{\xdef #1{(\number\nfig)}
\ifnum\nfig = 1\global\xdef\thefigs{\item{(\number\nfig)} #2\ }
\else
\global\xdef\oldfigs{\thefigs}
\global\xdef\thefigs{\oldfigs\vskip.1in\item{(\number\nfig)} #2\ }%
\fi%
\global\advance\nfig by 1 } 

\def\figurecaptions{\vfill\eject\section{Figure Captions}\thefigs}

\def\fig#1{\xdef #1{(\number\nfig)}
\global\advance\nfig by 1 } 


\newcount\cflag
\newcount\nequation
\global\nequation = 1
\def\eqlabel{(1)}

\def\nexteqno{\ifnum\cflag = 0
\global\advance\nequation by 1
\fi
\global\cflag = 0
\xdef\eqlabel{(\number\nequation)}}

\def\lasteqno{\global\advance\nequation by -1
\xdef\eqlabel{(\number\nequation)}}

\def\label#1{\xdef #1{(\number\nequation)}
\ifnum\dflag = 1
{\escapechar = -1
\xdef\draftname{\littlefont\string#1}}
\fi}

\def\clabel#1#2{\xdef\eqlabel{(\number\nequation #2)}
\global\cflag = 1
\xdef #1{\eqlabel}
\ifnum\dflag = 1
{\escapechar = -1
\xdef\draftname{\string#1}}
\fi}

\def\cclabel#1#2{\xdef\eqlabel{#2)}
\global\cflag = 1
\xdef #1{\eqlabel}
\ifnum\dflag = 1
{\escapechar = -1
\xdef\draftname{\string#1}}
\fi}


\def\eeq{}

\def\eqnn #1\eeq{$$ #1 $$}

\def\eq #1\eeq{
\ifnum\dflag = 0
{\xdef\draftname{\ }}
\fi 
$$ #1
\eqno{\eqlabel \rlap{\ \draftname}} $$
\nexteqno}



\def\eol{& \eqlabel \rlap{\ \draftname} \crcr
\nexteqno
\xdef\draftname{\ }}

\def\eeol{& \eqlabel \rlap{\ \draftname}
\nexteqno
\xdef\draftname{\ }}



\def\eqa #1\eeq{
\ifnum\dflag = 0
{\xdef\draftname{\ }}
\fi 
$$ \eqalignno{ #1 } $$
\global\cflag = 0}


\def\ie{{\it i.e.\/}}

\def\etal{{\it et.al.\/}}

\def\via{{\it via\/}}

\def\cf{{\it c.f.\/}}


\def\npb#1#2#3{{\it Nucl.\ Phys.} {\bf B#1} (19#2) #3}
\def\plb#1#2#3{{\it Phys.\ Lett.} {\bf #1B} (19#2) #3}

\def\prd#1#2#3{{\it Phys.\ Rev.} {\bf D#1} (19#2) #3}

\def\prl#1#2#3{{\it Phys.\ Rev.\ Lett.} {\bf #1} (19#2) #3}


\global\nulldelimiterspace = 0pt



\def\frac#1#2{{{#1} \over {#2}}\,}  
\def\hf{{1\over 2}}
\def\nth#1{{1\over #1}}


\def\Dsl{\hbox{/\kern-.6700em\it D}} 
\def\dsl{\hbox{/\kern-.5300em$\partial$}}
\def\pxpsl{\hbox{/\kern-.5600em$p$}}
\def\ssl{\hbox{/\kern-.5300em$s$}}
\def\epssl{\hbox{/\kern-.5100em$\epsilon$}}
\def\delsl{\hbox{/\kern-.6300em$\nabla$}}
\def\lxpsl{\hbox{/\kern-.4300em$l$}}
\def\elxpsl{\hbox{/\kern-.4500em$\ell$}}
\def\kxpsl{\hbox{/\kern-.5100em$k$}}
\def\qxpsl{\hbox{/\kern-.5000em$q$}}
\def\sla#1{\raise.15ex\hbox{$/$}\kern-.57em #1}
\def\Pl{\gamma_{\sss L}}
\def\Pr{\gamma_{\sss R}}



\def\roughly#1{\mathrel{\raise.3ex\hbox{$#1$\kern-.75em\lower1ex\hbox{$\sim$}}}}
\def\lsim{\roughly<}
\def\gsim{\roughly>}

\def\tw#1{\tilde{#1}}
\def\ol#1{\overline{#1}}



\def\bfl{{\bf l}}

\def\bfp{{\bf p}}



\def\Sca{{\cal A}}

\def\Scf{{\cal F}}

\def\Scl{{\cal L}}

\def\Scn{{\cal N}}

\def\Scq{{\cal Q}}


\def\ssf{{\sss F}}

\def\ssl{{\sss L}}

\def\ssn{{\sss N}}

\def\ssr{{\sss R}}
\def\ssS{{\sss S}}
\def\sst{{\sss T}}


\def\Re{{\rm Re\;}}


\def\bra#1{\langle #1 |}
\def\ket#1{| #1 \rangle}

\def\Avg#1{\left\langle #1 \right\rangle}



\def\hc{{\rm h.c.}}


\def\eV{{\rm \ eV}}

\def\MeV{{\rm \ MeV}}
\def\GeV{{\rm \ GeV}}


\let\nopictures=Y

\nopageonenumber
\baselineskip = 16pt
\barsoff


\def\bb{\beta\beta}
\def\bbtn{\bb_{2\nu}}
\def\bbm{\bb_{\varphi}}
\def\bbmm{\bb_{\varphi\varphi}}

\def\bbzn{\bb_{0\nu}}
\def\pf{p_{\sss F}}
\def\EF{E_{\sss F}}
\def\GF{G_{\sss F}}

\def\veps{\varepsilon}

\def\bk{\item{}}
\def\bb{\beta \beta}
\def\bbzn{\beta \beta_{0\nu}}

\def\cs#1{c_#1}
\def\sn#1{s_#1}

\def\cst{\cs{\theta}}
\def\csp{\cs{\varphi}}

\def\snt{\sn{\theta}}
\def\snp{\sn{\varphi}}
\def\mi{m_{\nu_i}}
\def\mj{m_{\nu_j}}
\def\ma{m_{\ssn_a}}
\def\wf{w_\ssf}
\def\wgt{w_{\sss GT}}
\def\geff{g_{\rm eff}}


\line{hep-ph/9412365 \hfil McGill-94/18, NEIP-94-010, UMD-PP-95-78}

\title
\centerline{Multi-Majoron Modes for}
\centerline{Neutrinoless Double-Beta Decay*}
\endtitle
\footnote{}{*  {\eightrm Research partially supported by the Swiss National
 Foundation, NSERC of Canada, FCAR du Qu\'ebec,
and by National Science Foundation grant no. PHY-9119745.}}
\authors
\centerline{P. Bamert${}^a$, C.P. Burgess${}^b$ and R.N. Mohapatra${}^c$}
\vskip .15in
\centerline{\it ${}^a$ Institut de Physique, Universit\'e de Neuch\^atel}
\centerline{\it 1 Rue A.L. Breguet, CH-2000 Neuch\^atel, Switzerland.}
\vskip .1in
\centerline{\it ${}^b$ Physics Department, McGill University}
\centerline{\it 3600 University St., Montr\'eal, Qu\'ebec,  Canada, H3A 2T8.}
\vskip .1in
\centerline{\it ${}^c$ Department of Physics, University of Maryland}
\centerline{\it College Park, Maryland, USA, 20742.}
\endauthors

\abstract
We construct two new classes of models for double beta decay, each
of which leads to an electron spectrum which differs from the
decays which are usually considered.  One of the classes has a spectrum
which has not been considered to date, and which is softer than the usual
two-neutrino decay of the Standard Model. We construct illustrative
models to show how other phenomenological bounds can be accommodated.
We typically find that, although all other bounds can be satisfied, the
predicted double-beta decay rate only in one class of
 these models is at best large enough to be
just detectable  in current experiments.
\endabstract


\vfill\eject
\section{Introduction}

\ref\icnapp{C.P. Burgess and J.M. Cline, in the proceedings of {\it The 1st
International Conference on Nonaccelerator Physics}, Bangalore, January 1994,
(World Scientific, Singapore).}
The purpose of this paper is to present two classes of double-beta
($\bb$) decay, which have not been hitherto considered by workers in the field,
and are potentially experimentally distinguishable from the presently-known
kinds of decays. Besides presenting a general discussion of the phenomenology
of
$\bb$ decay in the kinds of theories which we consider, we also explore the
 other
kinds of experiments to which these models might be expected to contribute. In
so doing we fill in the last missing items in the classification of $\bb$ decay
that was presented in ref.~\icnapp.

First, a brief motivation for this calculation.

\ref\expreview{The experimental situation has recently been thoroughly
reviewed in: M. Moe, {\it Int.\ J.\ Mod.\ Phys.} {\bf E2} (1993) 507;
M. Moe and P. Vogel, {\it Ann. Rev. of Nucl. and Part. Sc.} (to appear).}
The past ten years have seen a great deal of effort go into the experimental
search for $\bb$ decay, an extremely rare process in which
two neutrons simultaneously decay into protons with the accompanied
emission of two electrons. This process --- when it occurs together with the
emission of two antineutrinos ($\bbtn$) --- is predicted to occur at
second-order in the charged-current weak interactions in the Standard Model
(SM).  The experimental effort has borne fruit in recent years, with the
unambiguous  observation of this decay in a number of different nuclear
isotopes
\expreview.

\ref\GR{G.B. Gelmini and M. Roncadelli, \plb{99} {81}{411}.}
\ref\GGN{H.M. Georgi, S.L. Glashow  and S. Nussinov, \npb{193} {81)}{297}.}

Of course, part of the interest in these experiments was originally motivated
by the possibility of  observing other decays, besides the SM process. There
are several other types of decay modes to which these experiments are
sensitive, such as the lepton-number violating neutrinoless process ($\bbzn$),
or the decay ($\bbm$) into two electrons plus a neutral scalar, $\varphi$. Both
of these reactions have been predicted by a number of plausible alternatives to
the SM, starting with the simple Gelmini-Roncadelli (GR) model of lepton-number
breaking \GR, \GGN.

The experimental quantity that is used to distinguish such exotic decays from
the run-of-the-mill SM events is the shape of the decay rate as a function of
the energies, $\veps_1$ and $\veps_2$, of the two emitted electrons.
This shape is simplest for $\bbzn$ decay, since in this case the rate
is zero unless the sum of the two electron energies, $\veps = \veps_1 +
\veps_2$, takes on a specific value, $Q$. $Q$ represents the total energy
that can be carried off by the two electrons, and is given in terms of the
relevant masses --- denoted respectively by $M$ and $M'$ for the
parent and daughter nuclei --- by $Q = M - M'$. In all
of the decays of interest $Q \simeq 2$ MeV in size. Other decays  (like $\bbtn$
and $\bbm$), which involve other relativistic decay products in addition to the
final electrons, instead predict a continuous decay distribution throughout the
entire interval $2 m_e \le \veps \le Q$.

\figure\spectra{The $\bb$ spectrum as a function of the two electrons' total
kinetic energy for various choices of the `spectral index' $n$. $n=1$
corresponds to the dotted line, $n=3$ is the dashed line, $n=5$ is the
solid line and $n=7$ is the dash-dotted line. All four curves have been
arbitrarily assigned the same maximal value for purposes of comparison.}

The spectrum that is predicted by the various types of decays turns out, quite
generally, to take a particularly simple form that is completely characterized
by a  single integer, or `spectral index', $n$. Explicitly:
\label\spectrum
\eq
{d \Gamma \over d\veps_1 d\veps_2} = C (Q - \veps_1 -
\veps_2)^n \;  \left[ p_1 \veps_1 F(\veps_1) \right]  \; \left[ p_2
\veps_2 F(\veps_2) \right]  ,
\eeq
where $C$ is independent of $\veps_1$ and $\veps_2$, and $p_i = |\bfp_i|$,
for $i=1,2$, represents the magnitude of the three-momentum of the
corresponding  electron. The quantity $F(\veps_i)$ is the Fermi function which
describes the  spectral distortion due to the electric charge of the nucleus.
The spectral shape which follows from eq.~\spectrum\ is shown, for various
choices of $n$, in Fig.~\spectra.

It is ultimately the small size of the energy, $Q \sim 2$ MeV, in comparison
with the typical momenta, $\pf \sim 60$ MeV, of the nucleons in the decaying
nucleus that ensures that the electron spectrum takes such a simple form ---
\ie\ one that is characterized simply by the integer $n$. This is because the
small ratio of these two scales justifies keeping only the lowest term in an
expansion of the decay rate in powers of the momenta of the final electrons
(and/or scalars and neutrinos).
The index, $n$, that is appropriate for any particular kind of decay
is determined by the leading term in this expansion of the relevant decay rate.
For example, for the SM process, $\bbtn$,
the phase space volume element for the final neutrinos implies an index
$n(SM) = 5$. For the scalar decay, $\bbm$, of the GR model, on the other
hand, the phase space of the lone scalar implies $n(GR) = 1$.  The difference
between these predictions --- \cf\ Figure \spectra --- forms the basis for the
experimental discrimination of these two models.

\ref\CMM{C.P. Burgess and J.M. Cline, \plb{298} {93}{141}; \prd{49}{94}{5925}.}
\ref\BSV{Z.G. Berezhiani, A.Yu. Smirnov and J.W.F. Valle, \plb{291}{92}{99}.}
\ref\carone{C.D. Carone, \plb{308}{93}{85}.}
\ref\LEP{D. Schaile, in the proceedings of the XXVIIth International
Conference on High Energy Physics, Glasgow, July 1994.}
\ref\excess{F.T. Avignone III \etal, in {\it Neutrino Masses and Neutrino
Astrophysics,}  proceedings of the IV Telemark Conference, Ashland, Wisconsin,
1987,  edited by V. Barger, F. Halzen, M. Marshak and K. Olive   (World
Scientific, Sinagpore, 1987), p. 248; \bk
M. Moe, M. Nelson, M. Vient and S.  Elliott, {\it Nucl. Phys.}
(Proc. Suppl.) {\bf B31} (1993).}
\ref\gone{See, for example, M. Moe, ref. \expreview.}

Two developments have recently provoked a theoretical re-evaluation
\CMM, \BSV, \carone\ of the kinds of new physics to which the current
experiments can be sensitive. The first of these was the advent of
precision measurements of the properties of the $Z$ resonance at LEP \LEP,
which has ruled out the original GR-type models for new physics. The
other development has come from the $\bb$ experiments themselves,
starting with the appearance of several indications of an excess of electrons
just below the endpoint in some of the experimental data for several types
of decays \excess.  Although the evidence for the excess has since
diminished \gone, the conclusion reached by the theoretical re-evaluation
still stands: viable theories which can produce observable $\bb$ decay are
possible, and can be consistent with all other, non-$\bb$, experimental limits,
but only if the new physics has rather different properties than have
previously
been assumed based on experience with GR-type models.

One of the main surprises which arose from this theoretical re-examination
has been the observation that new physics can be very generally divided into a
small number of classes --- depending on how it addresses a few basic issues
--- and that the experimental features that any model predicts for $\bb$ decay
is a robust  indicator of the class to which it belongs \icnapp.  (These issues
and categories are summarized in Table I.\foot\error{The table presented here
differs slightly from the table in ref.~\icnapp. A trivial change is our
introduction of new categories ID and IE, even though these are in practice
indistinguishable from IIC. Also, the index listed here for category IIE
reflects the results of the present paper, and differs from the preliminary
guess for this value which was given in ref.~\icnapp.})

\midinsert
$$\vbox{\tabskip=0pt \offinterlineskip
\halign to \hsize{\strut \tabskip=0pt \hfil#\hfil & \hfil#\hfil &\hfil#\hfil
&\hfil#\hfil &\hfil#\hfil &\hfil#\hfil \cr
\noalign{\hrule}\noalign{\smallskip}\noalign{\hrule}\noalign{\medskip}
& $L_e$ & A New Scalar: & $\bbzn$ & Dominant & \ \ Spectral
\cr\noalign{\medskip}
&  &  &  & \ \ Scalar Decay & \ \ Index
\cr
\noalign{\medskip}\noalign{\hrule}\noalign{\smallskip}
\noalign{\hrule}\noalign{\medskip}
IA & Broken & Does Not Exist & Yes & None & N.A. \cr
IB & Broken & Is Not a Goldstone Boson & Yes & $\bbm$ & $n=1$ \cr
IC & Broken & Is a Goldstone Boson & Yes & $\bbm$ & $n=1$ \cr
ID & Broken & Is Not a Goldstone Boson & Yes & $\bbmm$ & $n=3$ \cr
IE & Broken & Is a Goldstone Boson & Yes & $\bbmm$ & $n=3$ \cr
\noalign{\medskip}\noalign{\hrule}\noalign{\medskip}
IIA & \ Unbroken & Does Not Exist & No & None & N.A. \cr
IIB & \ Unbroken & \ \ \ Is Not a Goldstone Boson ($L_e=-2$)\ \ \
& No & $\bbm$ & $n = 1$ \cr
IIC & \ Unbroken & \ \ \ Is Not a Goldstone Boson ($L_e=-1$)\ \ \
& No & $\bbmm$ & $n = 3$ \cr
IID & \ Unbroken & Is a Goldstone boson ($L_e=-2$) & No & $\bbm$ & $n=3$ \cr
IIE & \ Unbroken & Is a Goldstone boson ($L_e=-1$) & No & $\bbmm$ & $n=7$ \cr
\noalign{\medskip}\noalign{\hrule}\noalign{\smallskip}\noalign{\hrule}
}}$$
\centerline{{\bf Table I}}
\medskip
\centerline{ A list of alternatives for modelling double beta decay.}
\endinsert

As is argued in ref.~\icnapp, the observed absence of $\bbzn$ decays at the
endpoint of the electron spectrum strongly constrains the classes of models for
which lepton number is broken (\ie\ classes IA -- IE). In particular, it
implies
that all of the scalar-emitting decays in these classes are essentially
indistinguishable from their counterparts in the lepton-number preserving
classes. That is to say, so long as $\bbzn$ decays are not observed, models in
class IB and IC are at present experimentally indistinguishable from those in
class IIB. Similarly, classes ID and IE cannot be distinguished from class IIC.

\ref\twoscalar{R. Mohapatra and E. Takasugi,\plb{211}{88}{192}.}
\ref\kai{Kai Zuber, in Relativistic Astrophysics and Particle Cosmology, ed.
by C.W. Akerlof and M.A. Srednicki, Annals of the New York Academy of Sciences,
Vol. 688 (New York, 1993).}
\ref\expgeff{See e.g. M. Beck {\it et.al.}, \prl{70}{93}{2853} and
references therein, or J.-L. Vuilleumier {\it et.al.}, {\it Nucl.
Phys. (Proc. Suppl.)} {\bf B31}{ (1993)}.}
Interestingly, all of the models which had been considered previously,
as well as most of the more recent proposals,  fall into only a few of the
several possible categories --- cases IB and IC of Table I. The other,
unorthodox, categories remain essentially unexplored. The two exceptions to
this
statement are case IID, which was discovered and studied in ref.~\CMM, and case
IE, which contains a supersymmetric model \twoscalar, in which {\it two}
scalars
are emitted in $\bb$ decay: $\bbmm$. Unfortunately, this last model has since
been ruled out by the measurements at LEP.  Both of these new classes of models
predict a $\bbm$ spectrum having $n = 3$. The resulting shape is intermediate
between the usual SM and GR spectra, and can be distinguished from these in
current experiments \kai , \expgeff.

The purpose of the present paper is to explore in detail those classes of
theories of Table I which have not been considered to date. Keeping in mind
the indistinguishability of classes IB, IC and IIB, as well as ID, IE and
IIC, we see there are two types of models which we must here consider: those
of classes IIC and IIE, both of which involve $\bb$ decay accompanied by the
emission of {\it two}, rather than just one, scalar particles. Two scalars must
be emitted because of lepton-number conservation, and the charge assignments of
the light particles. Models in these two categories differ from one another
according to whether or not the light scalar particle that is emitted is a
Goldstone boson.

We first examine models of category
IIC to show that these can be altered to evade detection at LEP, and we then
work through the completely new category, IIE, which predicts a spectral index,
$n=7$, that is completely different from all previous cases. Since this index
is
{\it larger} than that for the SM decay --- $n(SM)=5$ --- the resulting
spectral
shape is {\it softer} than the presently-observed $\bbtn$ spectra. This kind
of softer spectrum is much more difficult to distinguish from the experimental
background, and so we expect the experimental limit for the branching ratio
into the new $n=7$ decay mode to be comparatively weak.

On the theoretical side, we find that the nuclear form factors which appear in
all of the new $\bbmm$ decays are well understood, since they are the same as
those that appear in the more well-known decay modes, such as for $\bbzn$ and
$\bbm$ in the GR model. We find that using these matrix elements to estimate
the
potential size of the $\bbmm$ decay rate typically gives a result which can be
close to the present experimental limit for the $n=3$ decays, provided we relax
the conditions that are cast on the relevant masses by big bang nucleosynthesis
constraints. The estimated rate for the $n=7$ decays appears to be too small to
be detectable.

This paper has the following organization. The next section computes the
$\bbmm$ decay rate for the two classes of decays (IIC and IIE). We do so
to verify their spectral index --- which is their experimental
signature --- and to determine how the resulting rates depend on the couplings
in underlying models. For class IIC (having $n=3$) we consider two types of
decay mechanisms: those which proceed due to the exchange of new sterile
neutrinos, and those which proceed through the interactions of new scalar
particles. Representative models for decays in class IIC are then
constructed in sections 3, and those for class IIE in section 4. Section 5
then summarizes the most important phenomenological bounds for these models.
Our conclusions are finally summarized in section 6.

\section{Two-Scalar $\bb$ Decays}

Before constructing models in detail, we first pause here to record
expressions for the decay rate into two electrons plus two scalars,
$\Gamma(\bbmm)$. We do so in order to display how the $\bbmm$
rate depends on the various masses and couplings of the theory. We
may then use this dependence to see what kinds of new particles can
be used to generate an observable signal in $\bb$ experiments. We
defer the discussion of any further phenomenology, which requires the
construction of an explicit model, to the following sections.

We consider here three kinds of scenarios. First we consider two types of
models for which $\bbmm$ decay is mediated by the admixture of ordinary
neutrinos with new sterile fermions. The two kinds of such decays which we
consider are those for which the scalar emission is either derivatively
suppressed (\ie\ case IIE, with $n=7$), or not (case IIC with $n=3$).
Finally we consider the alternative for which $\bbmm$ decay occurs because
of the interactions of new types of scalar particles. Such scalar-mediated
decays can also arise with or without derivative suppression for the
final scalar emission (classes IIE or IIC), but since the decay rate
for the class IIE models are generally very small, we do not pursue them
in detail for the scalar-mediated case.

\subsection{Fermion-Mediated $n=3$ Decays}

Consider first a theory of neutrinos, $\nu_i$
and $N_a$, which respectively carry lepton number  $L_e(\nu_i) = +1$
and $L_e(N_a) = 0$, and which are coupled to a scalar, $\phi$, which has
lepton number $L_e(\phi) = +1$. The most general renormalizable and
$L_e$-conserving Yukawa couplings involving these fields is:
\label\yukawa
\eq
\Scl_{\rm yuk} = - \, \ol{\nu}_i \left( A_{ia} \Pl + B_{ia} \Pr
\right) N_a \; \phi + \hc ,
\eeq
where $A_{ia}$ and $B_{ia}$ represent arbitrary Yukawa-coupling matrices, and
$\Pl$ and $\Pr$ denote the usual projections onto left- and right-handed
spinors. We use majorana spinors here to represent the neutrinos, so the
conjugate which appears in eq.~\yukawa\ is $\ol{\nu}_i = \nu_i^\sst \, C^{-1}$,
where $C$ is the charge-conjugation matrix.
Suppose also that a number of right-handed neutrinos are included in the
theory, in order to form generic lepton number conserving masses, $m_{\nu_i}$,
for the $L_e = 1$ states, and that the $L_e = 0$ neutrinos, $N_a$, have generic
majorana masses, $m_{\ssn_a}$.

\figure\decaygraph{The Feynman graph representing the leading order
contribution to the two-scalar $\bbmm$ decay through sterile lepton
exchange.}

As long as any of the $\nu_i$'s participate in the charged-current
weak interactions, the couplings of eq.~\yukawa\ generically produce $\bbmm$
decay due to the Feynman graph of Fig.~\decaygraph. (Two scalars must be
emitted
in this decay due to conservation of $L_e$.) Even though Fig.~\decaygraph\
arises at tree level, the four-momentum of the exchanged neutrino is not
determined by energy and momentum conservation. This is because the amplitude
for the decay of the two initial neutrons must be convoluted with the
wavefunction of the initial nucleus. The important momentum scale in the
integration over the exchanged neutrino is therefore set by the scale of
momenta
to which the nucleons within the initial and final nuclei have access. Since
this scale is much larger than the energy, $Q$, that is available for the final
electrons and scalars, it is a good approximation to neglect the outgoing
scalar
and electron energies in the decay amplitude. Evaluating the graph in this
approximation gives the following expression for the $\bbmm$ decay rate:
\label\ommdecayrate
\eq
d \Gamma(\bbmm) = {(\GF\cos\theta_c)^4 \over 8 (2\pi)^5} \; \left| \Sca(\bbmm)
\right|^2 \; (Q - \veps_1-\veps_2)^3 \prod_{k=1}^2 p_k \veps_k F(\veps_k) \;
d\veps_k,
\eeq
where $\GF$ is Fermi's constant, and $\theta_c$ is the Cabbibo angle.
The quantity $\Sca(\bbmm)$ represents the integral:
\label\matrixelement
\eq
\Sca(\bbmm) = \left( { 2 \over 3 \pi^2} \right)^{\hf} \sum_{ija} \int
{d^4 \ell \over (2\pi)^4} \; {V_{e\nu_i} V_{e\nu_j} \Scn_{ija} \; {W^\mu}_\mu
\over (\ell^2 + \mi^2 -i \epsilon) \, (\ell^2 + \mj^2 -i \epsilon) \,
(\ell^2 + \ma^2 -i \epsilon) } ,
\eeq
where $V_{e\nu_i}$ is the Kobayashi-Maskawa-type matrix for the leptonic
charged current. The factor, $\Scn_{ija}$, in the numerator of
eq.~\matrixelement\ denotes the following expression:
\label\numerator
\eq
\Scn_{ija} \equiv (-\ell^2) \Bigl[ A_{ia} B_{ja} \mi + A_{ja} B_{ia} \mj
+ B_{ia} B_{ja} \ma \Bigr] + A_{ia} A_{ja}  \mi \mj \ma .
\eeq

Finally, the tensor $W_{\mu\nu}$ is an independent function of the
space- and time-like parts, $\ell^0$ and $|\bfl|$, of $\ell_\mu$ in the rest
frame of the decaying nucleus, which encodes the nuclear matrix element that is
relevant to the decay \CMM:
\label\nmetensor
\eq
W_{\mu\nu} \equiv (2 \pi)^3 \, \sqrt{ { E E'\over M M'}} \;
\int d^4x \; \bra{N'} T^* \left[ J_\mu(x) J_\nu(0) \right]\ket{N} \;
e^{i\ell \cdot x} .
\eeq
Here $J_\mu = 2 \, \ol{u} \gamma_\mu \Pl d$ denotes the appropriate hadronic
charged current. The prefactor involving the energy-to-mass ratio --- $E/M$ and
$E'/M'$,  for the parent and daughter nuclei respectively --- is required in
order to ensure that $W_{\mu\nu}$ transforms under Lorentz transformations
as a tensor. The factor of $(2\pi)^3$ is purely conventional.

There are several features in these expressions that are noteworthy.

\item{1.}
Comparison of eqs.~\ommdecayrate\ and \spectrum\ shows
that the spectral index for this decay is $n=3$. The additional two powers
of $(Q-\veps)$, in comparison with the GR model, arise here simply due to
the additional phase space volume of the second scalar.

\ref\Halprin{A. Halprin, P. Minkowski, H. Primakoff and P. Rosen,
\prd{13}{76}{2567}.}
\ref\Haxton{W. Haxton and G. Stephenson, {\it Prog. in Particle and Nucl.
Physics} {\bf 12}, 409 (1984);
 J. Vergados, {\it Phys. Rep.} {\bf 133}, 1 (1986);
M. Doi, T. Kotani, H. Nishiura and E. Takasugi, {\it Prog. Theor. Phys.},
{\it Prog. Theor. Phys. Suppl.} {\bf 83}, 1 (1985).}
\ref\KK{H. Klapdor-Kleingrothaus, K. Muto and A. Staudt, {\it Europhys.
Lett.} {\bf 13}, 31 (1990); M. Hirsch et. al. {\it. Zeit. Phys. A}
{\bf 345}, 163 (1994);
H. Klapdor-Kleingrothaus , {\it Prog. Part. Nucl. Phys.} {\bf 32},
 261 (1994) for a review.}

\item{2.}
Eq.~\matrixelement\ depends on precisely the same combination of nuclear
form factors, ${W^\mu}_\mu = \wf - \wgt$, as do $\bbtn$ and $\bbzn$
decays, as well as $\bbm$ decay in GR-type models \CMM. (For future use we
express ${W^\mu}_\mu$ in terms of the Fermi and Gamow-Teller matrix elements,
$\wf = W_{00}$ and $\wgt = \sum_{i=1}^3 W_{ii}$. Ref.~\CMM\ gives $\wf$ and
$\wgt$ as explicit expressions in terms of the nuclear matrix elements
of various neutrino potentials.) Since these particular combinations of nuclear
matrix elements have been well studied within the context of these other
processes \Halprin, \Haxton, \KK, there is relatively little uncertainty in the
estimate for the total rate for this type of $\bbmm$ decay.

\item{3.}
The differential decay distribution for this decay as a function of the opening
angle, $\theta$, between the two electrons is exactly the same as it is
for $\bbtn$, $\bbzn$ and $\bbm$ decays (of both the GR type and
the new $\bbm$ decays of type IID). It is given explicitly (neglecting the
mutual repulsion of the outgoing electrons) by
\label\angulardist
\eq
{1 \over \Gamma} \; { d \Gamma \over d \cos \theta} = \hf \; \left( 1 -
v_1 v_2 \, \cos\theta \right) ,
\eeq
where $v_i = p_i/\veps_i$, for $i=1,2$, is the speed of the corresponding
electron in the nuclear rest frame.

\subsection{Fermion-Mediated $n=7$ Decays}

We next consider the case of double scalar emission, but in the special case
where the amplitude for emitting the scalars is suppressed by the scalar
energy.
This is the case when the scalars have purely derivative couplings, such as
when the scalar is a Goldstone boson which carries an unbroken lepton number
\CMM. For such models the expression given in eq.~\matrixelement\ for the
matrix
element, $\Sca$, vanishes identically, and so it is necessary to work to higher
order in the external scalar momenta in order to obtain a nonvanishing rate.
Since it turns out that the first nonvanishing contribution to the amplitude
arises at quadratic order in the scalar momenta, the calculation is quite
cumbersome when performed directly with the couplings of eq.~\yukawa.

A better way to proceed is to perform a change of variables so that the
derivative coupling nature of the Goldstone bosons is manifest from the
outset. Once this has been done, the trilinear coupling to neutrinos of a
Goldstone boson carrying $L_e = 1$, becomes:
\label\derivyukawa
\eq
\Scl_{\rm gb} = -i \; \ol{\nu}_i \gamma^\mu ( X_{ia} \Pl +
Y_{ia} \Pr) N_a \; \partial_\mu \phi + \hc,
\eeq
where the coefficients $X_{ia}$ and $Y_{ia}$ are computable \CMM\ in terms of
$A_{ia}$ and $B_{ia}$ of eq.~\yukawa, as well as the components of the
broken-symmetry generator, $\Scq$, for which $\phi$ is the Goldstone boson.

The contribution to $\bbmm$ decay in these models arises from the same Feynman
graph as before, Fig.~\decaygraph, but with the interaction of
eq.~\derivyukawa\ at each vertex. Evaluation of this graph gives a fairly
complicated expression for the decay rate, but this expression simplifies
greatly in  the special case of purely left-handed couplings, for which
$Y_{ia} = 0$. Since this case is sufficient to analyse the models we consider
in
later sections, we just quote here the decay rate in the limit $Y_{ia} = 0$:
\label\cmmdecayrate
\eq
d \Gamma(\bbmm) = {(\GF\cos\theta_c)^4 \over 8 (2\pi)^5} \; \left|
\tw{\Sca}(\bbmm) \right|^2 \; (Q - \veps_1-\veps_2)^7 \prod_{k=1}^2 p_k \veps_k
F(\veps_k) \;   d\veps_k,
\eeq
where the new quantity, $\tw{\Sca}(\bbmm)$, represents the integral:
\label\newmatrixelement
\eq
\tw{\Sca}(\bbmm) =  \left({ 4 \over 105 \pi^2} \right)^{\hf}
\; \sum_{ija} \int
{d^4 \ell \over (2\pi)^4} \; {V_{e\nu_i} V_{e\nu_j} \tw{\Scn}_{ija}
\; {W^\mu}_\mu \over (\ell^2 + \mi^2 -i \epsilon) \,
(\ell^2 + \mj^2 -i \epsilon) \, (\ell^2 + \ma^2 -i \epsilon) } .
\eeq
Apart from its normalization, eq.~\newmatrixelement, differs from
eq.~\matrixelement, obtained previously, only by the replacement $\Scn_{ija}
\to \tw{\Scn}_{ija}$, where:
\label\newnumerator
\eq
\tw{\Scn}_{ija} \equiv (-\ell^2)  X_{ia} X_{ja} \ma  .
\eeq

Once again there are several properties of the above expressions that bear
emphasis.

\item{1.}
The spectral index for this decay is $n=7$. The four additional powers
of $(Q-\veps)$, in comparison with eq.~\ommdecayrate, arise because each
scalar-emission vertex of eq.~\derivyukawa\ is explicitly proportional to
the scalar four-momentum.

\item{2.}
The decay rate depends only on the form factor combination
${W^\mu}_\mu = \wf - \wgt$, just as for the non-derivatively coupled
calculation. (This is by contrast with the situation for $\bbm$ decay, where
the
matrix elements for the derivatively-suppressed decay --- class IID --- differ
from those for the emission of a generic scalar --- class IIB.)

\item{3.}
The differential decay distribution as a function of the electron opening
angle remains exactly the same as for all of the other decays, as in
eq.~\angulardist.

\subsection{Scalar-Mediated $n=3$ Decays}

The previous calculations produced similar expressions for the decay amplitudes
--- \cf\ eqs.~\matrixelement\ and \newmatrixelement\ --- largely because
the same Feynman graph is responsible for it in both classes of models. In this
section we explore an alternative class of models for which $\bbmm$ decay
proceeds \via\ a different process: the exchange of an exotic set of scalars.

Consider therefore supplementing the SM by a collection of new scalars, which
we
classify using their lepton and electric-charge quantum numbers, $L_e$ and $Q$.
Besides the very light $(Q,L_e) = (0,+1)$ particle, $\varphi$, which is emitted
in the decay, we imagine also adding two other new classes of fields, $\chi_i$
and $\Delta_a$, which respectively have the $(Q,L_e)$ assignments: $(+1,-1)$
and
$(-2,+2)$. These quantum numbers permit the following types of $L_e$-invariant
trilinear-scalar, and Yukawa couplings:
\label\trilinears
\eqa
\Scl_{\rm tri} &= - \hf \, \mu^{ija} \, \chi_i \chi_j \Delta_a + \hc, \eol
\Scl_{\rm yuk} &= -  \hf \, \ol{e} (h_\ssl^a \Pl + h_\ssr^a \Pr) e^c
\, \Delta_a + \hc \, , \eeol
\eeq
in addition to the charged-current coupling
\label\scalarcc
\eq
\Scl_{\rm scc} = -i \lambda^i g W^\mu \Bigl( \chi_i \partial_\mu \varphi
- \partial_\mu \chi_i \, \varphi \Bigr) + \hc .
\eeq
Here $e^c$ denotes the Dirac conjugate of the electron field, and $\lambda^i$
parameterizes the strength of the scalar charged-current interaction relative
to
the usual $SU_\ssl(2)$ gauge coupling, $g$. For the sake of keeping our final
expressions simple, we assume from here on that the index `$a$' takes only one
value, so that there is only one field, $\Delta$, having the quantum numbers
$(Q,L_e) = (-2,+2)$. We therefore omit this index from the coupling
parameters as these are defined above: \ie| $\mu^{ija}\rightarrow\mu^{ij}$,
and $h^a_{\ssl ,\ssr}\rightarrow h_{\ssl ,\ssr}$. We also denote the mass
eigenvalues for $\chi_i$ and $\Delta$ respectively by $M_i$ and $M_\Delta$.

\figure\scalarfig{The Feynman graph contributing to $\bbmm$ decay purely due to
the self-interactions of the Higgs sector. }

With these couplings and interactions $\bbmm$ decay is mediated by the Feynman
graph of Fig.~\scalarfig. Evaluation of this graph gives the following decay
rate:
\label\scalarmedrate
\eq
d \Gamma_s(\bbmm) = {(\GF\cos\theta_c)^4 \over 8 (2\pi)^5} \; \left|
\Sca_s(\bbmm) \right|^2 \; (Q - \veps_1-\veps_2)^3 \prod_{k=1}^2
p_k \veps_k F(\veps_k) \; d\veps_k,
\eeq
where the scalar-mediated decay amplitude, $\Sca_s(\bbmm)$, is:
\label\scamatrixelement
\eq
\Sca_s(\bbmm) =  \left( {32 \over 3 \pi^2}\right)^{\hf} {h R \over
 M^2_\Delta } \;  \sum_{ij} \;
\int {d^4 \ell \over (2\pi)^4} \; {\lambda^i \lambda^j \mu^{ij}
\; [\wf \, \ell_0^2 + \nth{3} \, \wgt {\bf l}^2 ] \over
(\ell^2 + M_i^2 -i \epsilon) \, (\ell^2 + M_j^2 -i \epsilon) } \, .
\eeq
In this expression $h = \sqrt{|h_\ssl|^2 + |h_\ssr|^2}$, and
\label\rdef
\eq
R = \left( 1 +{\xi m_e^2 \over (h)^2 \veps_1 \veps_2} \right)^{\hf},
\eeq
with $\xi = 2 \Re[h_\ssl h_\ssr^*]$ and $m_e$ denoting the electron
mass.

These expressions differ from those for the fermion-mediated decays in a
number of interesting ways.

\item{1.}
The spectral index for this decay is determined by phase space, and is $n=3$.
This is because the assumed $\varphi$ couplings are not derivatively
suppressed.
A class of similar $n=7$ decays can be obtained by imposing the additional
requirement of derivative coupling. We do not consider such decays explicitly
here.

\item{2.}
Once more the decay rate depends only on the familiar form factors,
$\wf$ and $\wgt$, although this time they do not arise in the particular
linear combination ${W^\mu}_\mu = \wf - \wgt$.

\item{3.}
If the $\Delta$-electron coupling has a right-handed component, $h_\ssr \ne 0$,
then the differential decay distribution as a function of the electron opening
angle {\it differs} from that of all of the other decays we have considered.
Assuming only one species of $\Delta$, its explicit expression is,
\label\newangdist
\eq
{1 \over \Gamma} \; { d \Gamma \over d \cos \theta} = \hf \; \left( 1 -
{v_1 v_2 \over R^2} \, \cos\theta \right) ,
\eeq
so the deviation of the decay distribution from that of eq.~\angulardist\ is
controlled by the difference between $R$ and unity.

We next turn to the construction of some underlying models for each of
these types of decays.

\section{Model-Building: The Case of Spectral Index $n=3$.}

We first consider models for which the dominant contribution to $\bb$ decay
is through the emission of two scalars, but for which the emission of these
scalars is not suppressed by powers of the scalar momentum, {\it i.e.} models
of type IIC. As was mentioned earlier, and as is discussed in some detail in
refs.~\icnapp\ and \CMM, this also {\it includes} those models for which the
light scalar is the Goldstone boson for lepton number (class IE), even though
the couplings of such a scalar can always be put into the derivative form of
eq.~\derivyukawa.\foot\hereswhy{Briefly, with derivatively-coupled variables
the
dominant contribution to $\ss \bb$ decay for small scalar momentum comes from
graphs in which the scalar is emitted from the $\ss electron$ line, for which
an
infrared singularity compensates for the derivative coupling.}

In the following sections we consider a few representative models which produce
this kind of decay.

\vfill\eject
\subsection{Decays Mediated by Sterile Neutrinos}

\ref\CMP{Y. Chikashige,  R.N. Mohapatra and R.D. Peccei, \prl{45}{80}{1926}.}

A mechanism for producing $\bbmm$ decay, without running
into conflict with the LEP bounds, is to require the light scalar to be an
electroweak singlet, which couples to the usual electroweak neutrinos
through their mixing with various species of sterile neutrinos. This
results in a variation of the old singlet-majoron model \CMP.
In this way an unacceptable contribution to the invisible $Z$ width
can be avoided. A simple realization of this idea requires the introduction of
three species of electroweak-singlet, left-handed neutrinos: $N_\pm$ and
$N_0$, together with a single scalar field, $\phi$, carrying lepton
number +1. The subscript $i=0,\pm$ on $N_i$ indicates the lepton number of the
corresponding left-handed state.

Ignoring, for simplicity, all mixings with $\nu_\mu$ and $\nu_\tau$,
we are led to the following, most general, lepton-number-invariant
mass and Yukawa terms involving the new particles:
\label\myterms
\eq \eqalign{
\Scl_m &= - \, M (\ol{N}_+ \Pl N_-)  - {m \over 2} \,
(\ol{N}_0 \Pl N_0) + \hc , \cr
\Scl_y &= - \lambda (\ol{N}_- \Pl L) \; H - g_- \, (\ol{N}_- \Pl N_0) \; \phi -
g_+ \, (\ol{N}_+ \Pl N_0) \; \phi^* + \hc. \cr}
\eeq
Here $L = {\nu_e \choose e}$ and $H = {H^+ \choose H^0}$ represent
the usual SM lepton and Higgs doublets. (Where necessary the
conjugate Higgs doublet is denoted by $\tilde{H} = i\sigma_2 H^*$.)
Finally, recall that our use of majorana spinors implies $\ol{N}_i
= N_i^\sst C^{-1}$.

\ref\oscar{C.P. Burgess and O. Hern\'andez, \prd{48}{93}{4326}.}

Although, in general, both $H^0$ and $\phi$ could acquire vacuum
expectation values ({\it vev}'s): $\Avg{H^0} =v$ and $\Avg{\phi} = u$,
in practice it is a good approximation to work in the limit $u \approx 0$.
This is because the absence of an observable $\bbzn$ decay signal implies
the combination $g_i u$ must be small. On the other hand, the couplings
$g_i$ themselves cannot be too small or else the $\bbmm$ decay of
interest here will itself be unobservable. As a result the {\it vev}
$u$ must itself be constrained to be negligible in comparison to the
other masses of the problem. We therefore neglect $u$ in all of
what follows.\foot\natural{Notice that, unlike the choice $\ss u = 0$,
a very small but nonzero value for $\ss u$ is quite difficult to make
natural, even in the technical sense \CMM, \oscar.} We nevertheless
assume that the scalar, $\phi$, is light enough to permit $\bbmm$ decay
to occur. This can be done by tuning the mass in the potential to
be small, or by permitting $u \ne 0$ but with $u$ tuned to be small
enough to not produce too much $\bbzn$ decay. Either choice involves
a certain amount of arbitrariness.

The neutrino mass spectrum for this model greatly simplifies in the $u=0$
limit. There is one $L_e=0$ mass eigenstate, $N_0$, with mass $m$;
there is a massless state, $\nu'_e = \nu_e \cst - N_+ \snt$; and there is
a massive, Dirac neutrino, $\nu_s = \Pl N'_+ + \Pr N_-$, with
$N'_+ = N_+ \cst + \nu_e \snt$ and having mass $M_s =
\sqrt{M^2 + \lambda^2 v^2}$. Here $\cst$ and $\snt$ denote the cosine
and sine of the mixing angle, $\theta$, which is defined by: $\tan \theta =
\lambda v/M$.

The Yukawa couplings of these mass eigenstates take the form of eq.~\yukawa,
with
\label\explicityukawas
\eq
A_{\nu_e' \ssn_0} = 0, \qquad A_{\nu_s \ssn_0} = g_-, \qquad
B_{\nu_e' \ssn_0} = - g_+ \snt, \qquad B_{\nu_s \ssn_0} = g_+ \cst.
\eeq
Using these expressions in the general result, eq.~\matrixelement, then
gives:
\label\explicitmatrixelement
\eq
\Sca(\bbmm) = \left( { 2 \over 3 \pi^2} \right)^{\hf} \; M_s^2 \snt^2
\; \int {d^4 \ell \over (2\pi)^4} \; { {W^\mu}_\mu  ( g_-^2 m \, \ell^2 +
2 g_+ g_- \cst M_s \, \ell^2 - g_+^2 \cst^2 M_s^2 m ) \over
(\ell^2 -i \epsilon) \, (\ell^2 + M_s^2 -i \epsilon)^2 \, (\ell^2 + m^2 -i
\epsilon) } .
\eeq

This expression implies that the $\bbmm$ decay rate is maximized for
large couplings and mixing angles, and for sterile-neutrino masses
in the vicinity of the nuclear-physics scale, $\sim (10 - 100)$ MeV,
which defines the important integration region in
eq.~\explicitmatrixelement. In order to see how big this rate can get
we therefore have to see how close we can get to this optimal range using
phenomenologically acceptable values for the various parameters. The
relevant constraints are summarized in section 5, and
they suggest that the amplitude can be made biggest for the following
values of masses and mixing angles:
\label\maxpar
\eq M_s \sim 350 \;{\rm MeV} ;\;\; m\sim 1  \MeV;\;\;\theta
\sim7.5\times10^{-2};\;\;g_+\sim 0.2
\eeq
or, provided we relax the bound on $m$ coming from nucleosynthesis (as
explained in section 5)
\label\maxpartwo
\eq M_s\sim m \sim 350 \;{\rm MeV} ;\;\;\theta
\sim7.5\times10^{-2};\;\;g_+\sim 1
\eeq
Interestingly, the coupling constant $g_-$ need not be large. Notice also
that the smallness of $\theta $ implies $\lambda v \ll M_s \approx M$.

Using these values, and recalling that the integration momentum satisfies $\ell
\lsim 60 \, \MeV \ll M_s$, we find eq.~\explicitmatrixelement, to be given
approximately by
\label\simpleampl
\eq
\Sca(\bbmm) = \left( { 2 \over 3 \pi^2} \right)^{\hf} \;g_+^2 s_\theta
^2 \, c_\theta^2 \, m \int{{d^4\ell\over (2\pi )^4} \;{{W^\mu}_\mu\over (\ell^2
-i \epsilon ) (\ell^2+m^2 - i\epsilon)}}.
\eeq

For historical reasons, experimentalists conventionally state their results
concerning the nonobservation of a scalar-emitting decay in terms of the old
Gelmini-Roncadelli type $\bbm$ decay. The quantitative limit is expressed
in terms of an upper bound to a dimensionless effective coupling constant,
$\geff$, which parametrizes the electron-neutrino/majoron
interaction in the GR model:
\label\geffdef
\eq
\Scl_{\rm phen} = {i \geff \over 2} \, \bar{\nu }_e \gamma_5 \nu_e
\; \varphi .
\eeq
This interaction gives a $\bbm$ decay rate which can be obtained from
eq.~\ommdecayrate\ by performing the replacement: $\Sca(\bbmm) \,
(Q - \veps_1 - \veps_2)^3 \to  \Sca(GR) \, (Q - \veps_1 - \veps_2)$,
with
\label\GRdecayrate
\eq
\Sca(GR) = 4i \sqrt{2} \, \geff \, \int{{d^4\ell\over (2\pi )^4}
\;{{W^\mu}_\mu\over \ell^2 -i \epsilon }}.
\eeq
Comparing this prediction with the results of current experiments gives an
upper
bound $\geff\lsim 10^{-4}$ \expgeff.

\ref\heavysterile{P. Bamert, C.P. Burgess and R.N. Mohapatra,
preprint McGill-94/37, NEIP-94-007, UMD-PP-95-11; {\it Nucl. Phys.
B} (to appear), hep-ph/9408367.}

To obtain a rough estimate of the sensitivity of current experiments to the
$\bbmm$ signal predicted by the model, we compare the $\bbmm$ decay rate with
the $\bbm$ rate which would be just detectable for the GR model, given the
current experimental limit $\geff\lsim 10^{-4}$. For this kind of  comparison
the precision of a detailed matrix element analysis is unnecessary,  so we
follow Refs.~\oscar\ and \heavysterile\ and make the following two
approximations, which permit an analytical calculation of the various rates.

\item
1 To perform the integral over $\ell$, we approximate the nuclear matrix
elements by replacing the form factors, $\wf$ and $\wgt$ by step functions in
energy and momentum: $w_i \approx w_i^0 \,
 \Theta(\EF-l_0) \,
\Theta(\pf-|\bfl|)$.
  $\bbtn$ decay rates are reproduced if we take $|\wf^0 - \wgt^0| \sim 4
\MeV^{-1}$, and the present experimental sensitivity to an electron-neutrino
majorana mass in $\bbzn$ decay requires a nuclear Fermi momentum of $\pf \sim
60$ MeV. The Fermi energy then becomes $\EF = \pf^2/2m_\ssn \sim 2$ MeV.
For example, with these approximations the decay amplitude for the GR model
becomes $\Sca(GR) \approx [8 \sqrt{2}/(2 \pi)^3] \geff (\wf^0 - \wgt^0) \pf
\EF$.

\item
2 To perform the phase-space integrals, $P_n = \int d\veps_1 d\veps_2 \veps_1^2
\veps_2^2 (Q-\veps_1 - \veps_2)^n$, we neglect both the electron mass, $m_e$,
and the Fermi functions, $F(\veps_i)$. The integrals then become: $P_1
= Q^7/1,260$; $P_3 = Q^9/15,120$; $P_5 = Q^{11}/83,160$; and $P_7 =
Q^{13}/308,880$.

With these estimates, the effective equivalent GR coupling for which the $\bbm$
decay rate equals the $\bbmm$ decay rate is
\label\nondergeff
\eq
\geff (\bbmm) \sim (0.008) \; g_+^2 s_\theta^2 c_\theta^2 \;
\left({Q\over\pf} \right) \approx 10^{-8}\;\; (10^{-6}).
\eeq
With eq.\maxpar\ (respectively \maxpartwo ) in mind, we take $g_+ \sim 0.2$
(respectively $g_+ \sim 1$), $s_\theta \sim 0.1$, $Q \sim 2$ MeV, $m\sim 2\MeV$
(respectively $m\sim 350 \MeV $)  and $\pf \sim 60$ MeV in obtaining this
number. Although a coupling of $10^{-8}$ is probably hopeless to observe in
present experiments (which can detect $\geff \gsim 10^{-4}$), one at the level
of $10^{-6}$ is close enough to observability to warrant a more detailed
comparison of this type of model with the data.

\subsection{Decays Mediated by Virtual Scalars}

We next turn to models for which the $\bbmm$ decay is mediated by
the self couplings in the scalar sector, without the need for any extra heavy
sterile leptons. We argue in this section that $\bbmm$ decay in these models is
always constrained by LEP data to be too small to be observed. To see how this
works, consider an example of such a model in which the standard model is
extended by the inclusion of: ($i$) a second SM Higgs doublet, $\chi$, which
has
non-zero lepton number $L= -1$; ($ii$) a singlet Higgs boson field, $\phi$,
with
lepton number $L=+1$, and ($iii$) a weak isotriplet scalar field, $\Delta$,
with
$L~=~-2$. These charge assignments permit the couplings considered in the
previous section: the $\Delta$ Yukawa couplings contain a term of the form
$L~L~\Delta$, which couples it to the electron, and the Higgs potential
contains
the following terms which are relevant to $\bbmm$ decay:
\label\newHiggs
\eq
V_{\rm tri} = \mu_1 \, H^{\dagger} \chi \; \phi + {\mu_2 \over 2} \;
\chi^\sst \Delta^\dagger \chi  + \hc .
\eeq

Provided that the singlet field, $\phi$, is sufficiently light, the diagram of
Fig.~\scalarfig\ then induces $\bbmm$ decay. In this model the extra scalar
fields do not have nonzero {\it vev}'s so that lepton number remains a good
symmetry and the usual neutrinos remain massless, as in the SM.

Neglecting $\EF^2 \wf^0$ in comparison to $\pf^2 \wgt^0$, we find the
approximate form for the amplitude of eq.~\scamatrixelement:
\label\appscamat
\eq
\Sca_s(\bbmm) \sim {1 \over (2 \pi)^3}\; {2 \over 3} \sqrt{32\over 3\pi^2} \;
{s_\theta^2 h \mu_2 \wgt^0 \pf \EF \over M_\Delta^2} \; \Scf(\pf,M_\chi),
\eeq
where $s_\theta \propto \mu_1$ is the mixing angle which controls the strength
of the mixing between the light mass eigenstate, $\varphi$, and the singlet
field, $\phi$. $\Scf(\pf,M_\chi)$ is the function obtained by performing the
$\ell$ integration, which takes the values $\Scf \sim 1$ for $M_\chi \lsim \pf$
and $\Scf \sim (\pf / M_\chi)^4$ for $M_\chi \gg \pf$. With these expressions
we find:
\label\scalestimate
\eq
{\geff}_s (\bbmm) \approx (0.02) \; { |\wgt^0| \over |\wf^0 - \wgt^0|} \;
h s_\theta^2 \; \left( { \mu_2 Q \over M_\Delta^2} \right) \; \Scf(\pf,M_\chi).
\eeq

As usual, the rate is largest if all of the exotic scalars have couplings that
are as large as possible, and masses that are in the range of $(10 - 100)$
MeV. This range is particularly dangerous for the phenomenology of the present
model, however, since the new scalars couple to the photon and to the $Z$.
Any such particles having masses $\lsim 50$ GeV would contribute unacceptably
to $e^+e^-$ annihilation, and so are ruled out. The effective coupling,
$\geff$,
for scalar-mediated $\bbmm$ decay rate is therefore suppressed at least by the
factors $(Q/M_\Delta) \, (\pf/M_\chi)^4 \sim 10^{-16}$, and so is much too
small
to be detectable.

Since the suppression factor due to the large scalar masses is so devastatingly
small, this rules out observable scalar-meditated $\bbmm$ decays in a much
larger class of models than those we have directly considered. That is, one
might entertain models for which scalar-mediated $\bbmm$ decay does not involve
the $W$ boson at all. After all, Fig.~\scalarfig\ works equally well if the
longitudinal parts of the $W$ lines are replaced by yet another exotic scalar
which couples directly to quarks. In this case the nuclear matrix elements can
be larger since they can involve quark operators other than the electroweak
charged current. Also, the Yukawa coupling of such a scalar to quarks could
easily be of order $10^{-1}$ --- a value which is much larger than the
corresponding value of $10^{-4}$ for the SM Higgs. Nevertheless, none of
these enhancements can overcome the suppression due to large scalar masses.

\section{Model-Building: The Case of Spectral Index $n=7$}

We now turn to a model which produces $\bb$ decay with spectral
index $n=7$, {\it i.e.} a model of type IIE.
To do so we require a model in which lepton number
is unbroken, and which contains a Goldstone boson, $\varphi$, that carries
lepton number +1.  This can only happen if the theory has a nonabelian
flavour symmetry which acts on leptons, and which contains ordinary
lepton number as a generator.

The simplest case is to take the lepton-number symmetry group to be $G = SU(2)
\times U_{\ssl'}(1)$, with $G$ broken down to the $U_\ssl(1)$ of lepton number.
This can be done by working with a variation of the model of ref.~\CMM. We
therefore add the following electroweak-singlet, left-handed fermions to the
standard model: $N = {N_+ \choose N_0}$, $s_0$ and $s_-$, which transform under
the flavour symmetry, $G$, as: $N \sim \left({\bf 2}, \hf \right)$, $s_0
\sim \left({\bf 1}, 0 \right)$ and $s_- \sim \left({\bf 1}, -1 \right)$. If the
unbroken lepton number is chosen to be $L_e = T_3 + L'$, where $L'$ is the
generator of $U_{\ssl'}(1)$, then the subscripts of $N_+$, $N_0$, $s_0$ and
$s_-$ give the corresponding particle's lepton charge. In order to implement
the
symmetry breaking pattern $G \to U_\ssl(1)$, we also introduce the
electroweak-singlet scalar field, $\Phi = {\phi_+ \choose \phi_0}$, which
transforms under $G$ as: $\Phi \sim \left({\bf 2},\hf \right)$.

For this model the most general renormalizable $G$-invariant mass
and Yukawa interactions involving the new fields are
\label\newmyterms
\eq \eqalign{
\Scl_m &= - \, {M \over 2} \, (\ol{s}_0 \Pl s_0) + \hc , \cr
\Scl_y &= - \lambda (\ol{s}_- \Pl L) \; H - g_- \, (\ol{N}_i \Pl s_-) \;
\Phi_j \, \epsilon^{ij} + g_0 \, (\ol{N}_i \Pl s_0) \;
\tw{\Phi}_j  \, \epsilon^{ij} + \hc. \cr}
\eeq

If the scalar fields acquire the {\it vev}'s $\Avg{H^0} = v$ and
$\Avg{\phi_0} = u$ (which we take, for simplicity, to be real),
then $G$ breaks to $U_\ssl(1)$ as required. If we also neglect mixings with
$\nu_\mu$ and $\nu_\tau$, the following neutrino spectrum is produced. First,
there is one massless $L_e = +1$ state, $\nu_e' =  \nu_e \cst - N_+ \snt$,
where
the mixing angle, $\theta$, satisfies: $\tan \theta = \lambda v/g_- u$. Then,
there is a massive $L_e =+1$ Dirac state, $\nu_s = \Pl N_+' + \Pr s_-$, with
$N'_+ = N_+ \cst + \nu_e \snt$, and with mass $M_s =  \sqrt{ \lambda^2 v^2 +
g_-^2 u^2}$. There are also two majorana $L_e =0$ states, $\nu_\pm$, which
respectively have masses $M_\pm = \hf \left[ \sqrt{ M^2 + 4 g_0^2 u^2} \pm M
\right]$. These mass eigenstates are related to $N_0$ and $s_0$ by:
\label\rotn
\eq
{\nu_- \choose \nu_+} = \pmatrix{ i\csp & i\snp \cr - \snp & \csp \cr}
\; { N_0 \choose s_0},
\eeq
with $\tan(2\varphi) = 2 g_0 u/M$. The factors of $i$ in this mixing matrix
are due to the chiral rotation needed to ensure that both mass eigenvalues,
$M_\pm$, are positive.

The next step is to rotate fields to make the derivative couplings of
the Goldstone bosons explicit (this is done in detail for a related
model in ref.~\CMM). The appearance of the derivative coupling can be seen as
follows: use exponential parameterization of the scalar fields that have
non-zero {\it vev} so that the Goldstone field appears in the exponential.
Then redefine the fermion fields so that the new fermion field is given by
$F^{\prime}\equiv F e^{i\varphi/v}$ (in this case $F^{\prime}$ is
the heavy neutral lepton). After this redefinition, the Yukawa interaction
becomes completely independent of the Goldstone field, which reappears with
a derivative coupling due to the fermion kinetic energy term. The final form
of the interaction given in eq.(8) then emerges on diagonalization of the
fermion mass terms with coupling matrices $X_{ia}$
and $Y_{ia}$ given by $Y_{ia} = 0 $, and
\label\xterms
\eq
X_{\nu_e' \nu_-} = i \; {\snt \csp \over u} , \qquad X_{\nu_e' \nu_+}
= \; {\snt \snp \over u}, \qquad X_{\nu_s \nu_-} = -i\;
{\cst \csp \over u},
\qquad X_{\nu_s \nu_+} = -{\cst \snp \over u}.
\eeq

With these expressions we may evaluate the $\bbmm$ matrix element
of eq.~\newmatrixelement. It becomes:
\label\newresult
\eq \eqalign{
\tw{\Sca}(\bbmm) &\;=  \;\;{ 2 \over \sqrt{105} \; \pi} \; {\snt^2 \cst^2
M_s^4 \over u^2}\cr
&\times\int {d^4 \ell \over  (2\pi)^4} \; { {W^\mu}_\mu \over (\ell^2 -i
 \epsilon) \,
(\ell^2 + M_s^2 -i \epsilon)^2 } \; \left[ {\snp^2 M_+ \over
(\ell^2 + M_+^2 -i \epsilon) }- {\csp^2 M_- \over (\ell^2 + M_-^2 -i
\epsilon) } \right] .\cr }
\eeq

We quantify the size of this rate in a similar way as already done for the
model of section 3.1, namely by means of estimating the magnitude of $\geff$.
We therefore choose values for masses and mixing angles that are compatible
with the phenomenological bounds of section 5, but which also yield a maximal
rate. Specifically we take
\label\maxparder
\eq
M_s \gsim 350 \;{\rm MeV} ;\;\; M_+,M_-\sim {\rm\ a\ few\ MeV};\;\;\theta
\sim7.5\times10^{-2};\;\;g_-\sim 0.2.
\eeq
The amplitude can then be written in the following approximate form:
\label\simpleampder
\eq \eqalign{
\tw{\Sca}(\bbmm) \;= &\;\; { 2 \over \sqrt{105} \; \pi} \; {\snt^2 \cst^2
g_-^2 M_+M_- \over M_s^2} \; (\snp^2 M_- -\csp^2 M_+)\cr
& \qquad \times \int {d^4 \ell \over  (2\pi)^4} \; { {W^\mu}_\mu \over (\ell^2
-i \epsilon) \, (\ell^2 + M_+^2 -i \epsilon)(\ell^2 + M_-^2 -i \epsilon)}.\cr }
\eeq
To see this keep in mind that
in the limit of small $\theta$ we have $u\sim{M_s\over g_-}$.
Comparing this with the amplitude of the GR model gives an equivalent
$\geff$ of
\label\geffder
\eq \geff \sim (0.0006) \, \snt^2 \cst^2 g_-^2 \left({Q^3 \over
M_s^2 \pf }\right) \sim 10^{-14}.
\eeq
where the values of eq.\maxparder\ have been applied and we take the mixing
angle to satisfy $\tan{(\varphi )}\sim 1/\sqrt{2}$ to maximize the rate
(\ie\ equivalently $M_+\sim2M_-$). Specifically we take $M_+\sim \EF \sim Q
\sim
2$ MeV, $g_-\sim 0.2$, $\snt\sim 0.1$ and $M_s \sim 6\pf \sim 360$ MeV.

In the above we have chosen 0.2 as the maximal value for $g_-$, as is
naively required by kaon decay measurements (see section 5). It is probable
that values as large as $g_- \simeq 1$ would actually prove to be consistent
with these measurements in a more careful treatment, however. This is because
the scalar's derivative coupling provides an additional suppression to its
contribution to kaon decays. If so, then the upper bound on $\geff$ can be
relaxed to around $\geff\lsim 10^{-12}$.

It is clear that this rate is far too small to have any chance of being
detected, largely due to the suppression by powers of $Q$ that originate with
the extremely soft electron spectrum. We consider this model therefore only as
an
example that such an exotic decay spectrum, with spectral index $n=7$, can {\it
in principle} exist. Given that this is also the kind of spectrum that is least
well constrained experimentally, since cuts usually exclude the lowest-energy
electrons which are usually the most contaminated by backgrounds, it would be
very interesting to construct a model which predicts a detectable rate. We have
so far been unable to find such a model.

\section{Phenomenological Constraints}

The previous sections have seen us use particular kinds of masses and couplings
in order to maximize the $\bbmm$ decay rate. The present section is meant to
justify the masses and couplings we have chosen. We do so by listing the main
phenomenological constraints which the models we consider must confront. Since
the models with the largest $\bbmm$ signals involve sterile neutrinos coupled
to light scalars, we focus principally on the constraints on these.

\subsection{Laboratory Limits}

\ref\Gilman{For a review on the different types of experiments constraining
heavy isosinglet leptons see e.g. F.J. Gilman, {\it Comments\ Nucl.\ Part.\
Phys.\ }{\bf 16 }{(1986) 231} and references therein.}

In most of the models considered in this paper the SM has been supplemented by
a
number of electroweak-singlet fermion and scalar fields. Of these, it is only
the mixing of the sterile neutrinos with (in this case) the electron neutrino,
$\nu_e$, that is bounded experimentally. Since the sterile neutrinos in the
models we consider dominantly decay into lighter neutrinos and scalars, all of
whom are invisible to present day detectors, the only bounds which need be
considered are those which do not assume a decay into a visible mode.
The bounds that do apply can be classified into two subclasses according to
whether they are due to low- or high-energy experiments \Gilman.

\ref\brit{D.I. Britton \etal, \prd{46}{92}{R885}; \prl{68}{92}{3000}.}
\ref\kaonbound{ R. Shrock, \plb{96}{80}{159};
T. Yamazaki \etal, in Proc. XIth Intern. Conf. on Neutrino Physics and
Astrophysics, K. Kleinknecht and E.A. Paschos (World Scientific, Singapore,
1984) p. 183.}

At low energies we have those experiments which examine the decays of pions
\brit\ and kaons \kaonbound\ at rest, and search for nonstandard contributions
to the decay rate and the outgoing electron spectrum. These experiments
constrain the masses of sterile neutrinos in a model-independent way from a few
MeV almost up to $\sim  100$ MeV for pion decays, and up to $\sim 350$ MeV for
kaon decays. Mixing angles, $U_{e i}$, with the electron neutrino, $\nu_e$, are
strongly constrained for sterile neutrinos with masses in this range. For
example, upper bounds on $|U_{ei}|$ can be as low as $7\times10^{-4}$ for a
mass
of $50$ MeV \brit.

\ref\scalarlimits{ V. Barger, W. Y. Keung and S. Pakvasa, \prd{25}{82}{907};
\bk T. Goldman, E. Kolb and G. Stephenson, \prd{26}{82}{2503}; \bk
A. Santamaria, J. Bernabeu and A. Pich, \prd{36}{87}{1408}; \bk
C. E. Picciotto {\it et.al.}, \prd{37}{88}{1131}.}

These same experiments, as well as measurements of the muon lifetime, also
limit
the possibility for the existence of weak decays into very light scalars in
addition to neutrinos \scalarlimits. Specifically these limits come from
precision measurements of the Michel parameter, $\rho$, in muon decay, and from
using the measured decay spectra of the outgoing leptons to constrain decays
such as $K \to \ell N \varphi$ or $\pi \to \ell N \varphi$. The failure to
observe such decays can be expressed as an upper bound on a hypothetical
dimensionless Yukawa coupling, $\tw{g}_{\rm eff}$, of an effective
$\nu_e-N-\varphi $ interaction. In terms of this coupling, the current bounds
are respectively $\tw{g}_{\rm eff} \lsim 5.7\times 10^{-2}$ (Muon decay),
$\lsim 1.6\times 10^{-2}$ (Pion decay) and $\lsim 1.3\times 10^{-2}$ (Kaon
decay) \scalarlimits.\foot\declimits{These bounds can be even more stringent if
additional model-dependent assumptions for the couplings involved are made.} Of
course, these bounds assume that the kinematics permit the emission of $N
\varphi$ in these decays. For the models of interest here, the bound therefore
applies if the $L_e= 0$ sterile fermions, $N_a$, are sufficiently light
compared
to the $\pi$, $K$ and/or $\mu$, as is certainly the case for $m_{\ssn_a} \sim
1$
MeV. For $m_{\ssn_a}$ in this mass range, the effective coupling that is then
bounded in this analysis turns out to be $\tw{g}_{\rm eff} \approx \snt \cst
g_+$. As a result, keeping in mind the bound on $\theta$ discussed below, $g_+$
must be smaller than $0.18$.

It is clear however that provided $N_a$ is too heavy to be produced in these
kinds  of decays (i.e. $m_{\ssn_a}\gsim 350\MeV$) then the bound described
above
no longer applies, and $g_+$ can be of order unity. We are led to consider
$m_{\ssn_a}\sim1 \MeV$  because of the specific way we choose to avoid conflict
with nucleosynthesis (see section 5.2). If another way to evade this bound
should be found, or if we simply accept the comparatively large number of
effective degrees of freedom contributed at this epoch, then $g_+$ can indeed
be
as large as $g_+\sim 1$.

For masses heavier than 350 MeV the best bounds on sterile neutrinos usually
come from beam dump experiments. However this type of experiment relies
crucially on the assumption that the isosinglet leptons decay via their charged
current weak interactions. The same is also true for bounds coming from
searches
on the $Z$-pole for exotic decays of the $Z$ boson, such as $Z\rightarrow
N\bar{\nu}\rightarrow  W^* e \bar{\nu}$. (Here $N$ denotes the isosinglet
state, and $W^*$ denotes a virtual $W$ boson.) However, since such decays do
{\it not} dominate in the models we consider --- the sterile states instead
prefer to decay into light scalars --- these experiments need not be considered
further here.

\ref\bigfit{W. Marciano and A. Sirlin, \prl{71}{93}{3629}; \bk
C.P. Burgess, S. Godfrey, H. K\"onig, D. London and I. Maksymyk,
\prd{49}{94}{6115}.}

The best limits, for the mass range above $\sim 350$ MeV, therefore come from
the reduction of the effective charged-current coupling of the electron, which
is suppressed by the cosine of the mixing angle between $\nu_e$ and the heavy
sterile state. This reduction potentially affects electroweak precision
experiments by influencing the measured value of Fermi's constant, $\GF$,
that is inferred from muon decay. It also shows up as a failure of lepton
universality in low-energy weak decays. This leads to the bound $|U_{ei}| <
7.5\times 10^{-2}$ ($2\sigma$) for masses above $\sim 350$ MeV \bigfit.
Another bound comes from the reduction of the neutral current coupling of the
electron-neutrino, leading to a decreased number of light neutrino species
as deduced from $Z$'s invisible width. However this bound is somewhat weaker
than the aforementioned universality bound, and applies only to sterile
neutrinos that are heavier than about $90 \GeV$, hence it is of no particular
importance in our case.

These bounds have immediate implications for the models considered here, such
as those discussed in sections 3.1 and 4. Since these models have only one
sterile  state, the heavy Dirac neutrino $\nu_s$, that mixes with $\nu_e$, we
need only  constrain the masses and mixings of this one particle. To maximize
the $\bbmm$ rate we prefer a mixing  angle, $\theta $, that is as large as
possible, and so we choose $\nu_s$ to be heavy enough to evade the bounds from
pion and kaon decay ---  $M_s \gsim 350\;$MeV --- and we take $\theta $ to
saturate the  universality bound --- $\theta \lsim 7.5\times 10^{-2}$.

\subsection{Nucleosynthesis}

\ref\BBN{T. Walker \etal, {\it Ap. J.} {\bf 376}, 51(1991). }
\ref\schramm{C. Copi, D. Schramm and M. Turner, preprint
FERMILAB-Pub-94/174-A, [astro-ph/9407006], (1994).}

Standard big-bang cosmology very successfully explains the abundances of light
elements produced during the nucleosynthesis epoch \BBN. In fact, this success
leaves little room for the existence of new species of particles at this
critical time. Specifically if there were new particles present at a
temperature of  $\sim 0.1 - 2$ MeV  their additional degrees of freedom would
increase the expansion  rate of the universe leading to an earlier freeze-out
of
neutron-proton  converting interactions and thus to a larger amount of neutrons
that  eventually would be cooked into ${}^4He$. The bound on the number
of additional degrees of freedom is conventionally stated in terms
of the maximal number of additional light neutrino species allowed
at nucleosynthesis $\delta N_\nu$ \schramm:
\label\deltanu
\eq
\delta N_\nu \lsim 0.4
\eeq

There are two simple ways to satisfy this bound when additional particle
types are present in a theory. For particles that are heavier than
about 1 MeV, the most straightforward way is to have them not be present
at all during nucleosynthesis. This can be achieved by either having them
decouple and subsequently decay sufficiently early --- keeping in mind the
possible contributions of the decay products --- or by having them remain in
thermal equilibrium as they become nonrelativistic, so that their abundance
becomes suppressed by the Boltzmann factor, $e^{-m/T}$. For particles which
are much lighter than 1 MeV, on the other hand, the contribution to $\delta
N_\nu$ is strongly suppressed if they decouple before the $QCD$ phase
transition
at $T\sim 200$ MeV. In this case they become sufficiently diluted compared to
ordinary particles afterwards due to the reheating of the photon bath during
this phase transition.

\ref\kainulainen{K. Enqvist, K. Kainulainen and M. Thomson,
\prl{68}{92}{744}; \npb{373}{92}{498}.}

Unfortunately, these methods alone are not enough to avoid conflict
of the models considered in this paper with nucleosynthesis. This is
because the very light scalars, that are present in all of our
models, couple through dimensionless couplings to ordinary neutrinos, such
as $\nu_e$. As a result, they remain in equilibrium with these neutrinos down
to temperatures that are well below 1 MeV. This makes a conflict with
nucleosynthesis generic for models which can produce $\bbm$ and $\bbmm$ decays.

Rather than therefore considering these models to be inevitably ruled out, we
demonstrate here an existence proof that the nucleosynthesis bound can be
avoided, given suitable masses for the sterile neutrinos. We propose this as an
example of the kinds of information that could be inferred from cosmological
bounds, should one of the exotic $\bb$ decays we describe ever be detected
experimentally. The loophole we describe follows its similar application in
ref.~\oscar, and is based on an observation of ref.~\kainulainen.

Suppose, then, that a heavy neutrino state either decays \kainulainen\ or
annihilates \oscar, into particles that are in equilibrium with ordinary
neutrinos, right after these neutrinos freeze out of chemical equilibrium with
the photon bath (\ie\ $T_\nu \sim 2.3$ MeV.) Provided this happens before the
neutron-proton ratio freezes out (at $T_{n/p}\sim 0.7$ MeV), then the abundance
of neutrinos will be increased relative to the standard cosmological scenario.
This overabundance of neutrinos will act to suppress the neutron abundance at
freezeout, and so decreases the  predicted production of ${}^4He$. As a result
it contributes to nucleosynthesis as a {\it negative} contribution to $\delta
N_\nu$ \kainulainen.

Consider, therefore, a single majorana neutrino with a mass in the critical
region (\ie\ $\sim 1-2$  MeV) which is still in equilibrium with $n_\ssS$
species of real scalars in addition to the $n_\ssf = 3$ species of ordinary
neutrinos, when it becomes nonrelativistic. Such a particle will annihilate
out right during the critical epoch, thereby heating the neutrino-scalar bath
above that of the photon bath. Then the magnitude of $\delta N_\nu$ that
results due to the above mechanism turns out to be \kainulainen, \oscar\
$\delta N_\nu = -4.6 \,\delta n$, where
\label\negdeltan
\eq
\delta n = \left( {n\over n_o}\right) \left( {E\over E_o} \right)^2 -1 = \left(
{{(n_\ssf + 1) + {4\over 7} \, n_\ssS}\over {n_\ssf + {4\over 7} \, n_\ssS}}
\right)^{5/3} -1.
\eeq
Here $n/n_o$ and $E/E_o$ denote the ratios of the electron-neutrino number
densities and energies after and before annihilation. The second
expression is obtained from the first one using entropy conservation.

We can now see how the models described in previous sections can accommodate
the nucleosynthesis bound. Take the model of section 3.1 as a representative
case. This model contains a massive Dirac state, $\nu_s$, which mixes
appreciably with the electron neutrino, and so which must be heavier than $\sim
350$ MeV to avoid the laboratory bounds described earlier. It also contains a
massive majorana neutrino, $N_0$, as well as a very light complex scalar.

Since the sterile neutrino is kept in thermal equilibrium with the photon bath
by neutrino scattering {\it via}  scalar ($\phi$) exchange, it remains in
equilibrium long after the temperature has dropped below its mass. Its
abundance
is therefore sufficiently suppressed to ensure that it never plays a role at
nucleosynthesis. The same would be true for the majorana neutrino if it were
heavy, however this would leave the light scalars in equilibrium at
nucleosynthesis, contributing $\delta N_\nu = {8 \over 7}$. The contribution of
these scalars can be very efficiently cancelled, however, if $N_0$ should have
a mass of around 1 MeV, since in this case it annihilates right in time to
reheat the neutrino sector.\foot\annihil{$\ss N_0$ remains in equilibrium
with $\ss \nu_e$ through scalar exchange.} In this case we can employ
eq.~\negdeltan\ with $n_F=3$ and $n_S=2$, yielding $\delta n = 0.43$ and so
a total shift in the effective number of neutrino species of $(\delta
N_\nu)_{\rm tot} = {8 \over 7} - 4.6 (0.43) = - 0.86$. Since $(\delta
N_\nu)_{\rm tot}$ rises to ${8\over 7}$ as $N_0$'s mass increases above
$T_\nu\sim 2.3$ MeV, it is possible to find an effective number of light
neutrinos which is compatible with observations. To summarize this paragraph:
Conflict with  nucleosynthesis can be avoided if $N_0$ has a mass in the range
of a few MeV.

Following similar lines of reasoning one can also make the model of section 4
satisfy the nucleosynthesis bound. In this case the annihilation of just one
majorana neutrino during the critical epoch is not enough though because
we have twice as many light scalar degrees of freedom as in the model discussed
above. Assuming however that the two majorana mass eigenstates $\nu_\pm$ are
both not much heavier than a few $\MeV$ solves the problem. To make this point
explicit we just state that if {\it both} $\nu_+$ and $\nu_-$ annihilated out
right in the critical epoch then this would result in $(\delta N_\nu)_{\rm tot}
= - 1$.

\subsection{Other Constraints from Astrophysics and Cosmology}

Other astrophysical bounds can constrain the properties of very weakly
interacting particles, such as the sterile neutrinos and scalars we are
entertaining. These turn out to not significantly constrain the models we
consider. We illustrate this point in this section by examining these bounds
for
the model considered in section 3.1.

\topic{(1) Stellar Evolution}
Sterile particles produced in the core of stars can disrupt our
understanding of stellar evolution since they can provide an extremely
efficient additional cooling  mechanism for a star. For the model of section
3.1, the sterile neutrinos $\nu_s$ and $N_0$ are simply too heavy to be
produced
in significant numbers inside the core of a star, since their masses are
respectively taken to be $\sim 350$ MeV and $\sim 1$ MeV. They are therefore
not
constrained by stellar evolution. The same conclusion also holds for the light
scalars of this model, although for different reasons. In this case the scalars
are light enough to be produced, but they only couple appreciably to neutrinos.
But since temperatures and densities inside a star are not sufficiently high to
produce a thermal or degenerate population of neutrinos, scalars are not
produced in significant quantities from the interior of the star. Direct
couplings with other particles, such as electrons, do arise due to loop
effects, but these are too small to be significant.\foot\doublycmm{See, for
example, the second Ref. of \CMM, where this issue is treated in more detail in
a similar context.}

\topic{(2) Supernovae}
Populations of neutrinos {\it are} maintained, however, within the cores of
supernovae, due to the much higher temperatures ($\sim 50-100$ MeV) and
densities that are found there.  Both the scalar, $\varphi $, and the sterile
majorana neutrino, $N_0$, are therefore sufficiently light to be produced in
significant numbers. However since both of these particles are also in thermal
equilibrium with the ordinary $SM$ neutrinos, they are trapped in the core of
the supernova and so do not lead to premature cooling.

\topic{(3) `Present-Day' Cosmology}
One last issue is the effect of $\varphi$ on structure formation, and on
the energy density of the universe observed today. Since $\varphi$ would remain
in  thermal equilibrium with the electron neutrinos (through exchange of
virtual
$N_0$) until a temperature in the region of $1 \eV$ if it was light, it
inevitably annihilates out to the neutrino sector once it becomes
nonrelativistic. It will therefore never dominate the energy density of the
universe and hence also not affect structure formation nor the age of the
universe in any significant way. This in turn means that $\varphi $ can have
any
mass below 1 $\MeV$.
\endtopic

\section{Conclusions}

In this paper we investigate a novel class of neutrinoless double-beta
decays, in which two (rather than one or no) scalars are emitted in addition to
the two observed electrons. We do so to provide sample models of the two
remaining classes of spectra for $\bb$ decays, which had not been hitherto
explored. We regard it as useful to construct representative models for each
class, since such models are required in order to determine the implications of
other experiments, at different energies, for $\bb$ decay.

We have obtained the following results:

\topic{1} We have found that two-scalar decays always lead to a $\bb$ electron
sum-energy spectrum whose spectral index is either $n=3$ or $n=7$. The $n=3$
spectrum is intermediate between the SM shape (for which $n=5$), and the
traditional scalar-emitting shape ($n=1$), such as is found in the
Gelmini-Roncadelli model. The $n=3$ spectrum is identical to a
recently-discovered class of models \CMM\ for which only a single scalar is
emitted during the decay.

\topic{2} The $n=7$ spectrum is completely new, and is {\it softer} than
the observed SM $\bbtn$ spectrum. As a result, it is more difficult to
experimentally extract, since most backgrounds tend to dominantly produce
low-energy electrons. It is intriguing that such an exotic spectrum could in
principle be hiding undetected even now in the present $\bbtn$ data!

\topic{3} Viable models can be constructed which predict an $n=3$ $\bb$ decay
rate which is reasonably close to the present experimental limit. In the viable
models with detectable rates the decay is mediated by the exchange of a number
of sterile neutrinos, whose masses must lie in the (1-100) MeV range. Their
couplings must also be as large as is permitted. The requirement of such large
couplings to obtain observable signals suggests that an observed $\bbmm$ signal
should be accompanied by the appearance of new physics in other experiments,
such as in violations of lepton universality.

A potential problem with this model is that it predicts an excess number of
degrees of freedom at nucleosynthesis that is equivalent to ${8 \over 7}$
of a neutrino, in comparison with the presently-quoted bound of $0.4$. A way of
evading this nucleosynthesis bound is described in section 5, but the required
couplings and masses for this scenario would predict a $\bbmm$ rate which is
two
orders of magnitude smaller, and so out of reach of current experiments.  We do
not regard this as sufficiently worrying to preclude searching for this decay
in
current experiments, since an experimental signal would likely prompt more
imaginative approaches to understanding nucleosynthesis in these models.

\topic{4} Because of its soft spectrum, we were unable to construct
viable models of the $n=7$ $\bbmm$ decay, for which the predicted rate is large
enough to be detectable. Thus, although the existence of an $n=7$ decay mode is
experimentally tantalizing, it is unlikely to be discovered in $\bb$ decay
experiments for the foreseeable future.

\topic{5} For the purposes of comparison we also examined a class of models
for which $\bbmm$ arises due to the mutual interactions amongst new particles
in
the scalar sector, rather than from new sterile leptons. Although the decay
rate that is predicted for these models is also too small to be seen, the
decays nevertheless exhibit novel features. One such is a nonstandard angular
distribution of the two electrons as a function of the opening angle between
them, similar to what arises in models with right-handed currents.
\endtopic

All of these results serve to underline the theoretical understanding which has
emerged over the past few years. Scalar-emitting $\bb$ decays can
reasonably be expected to be observed, although the properties of the
phenomenologically viable models which can do so are very different than what
would be expected based on intuition that is based on the original GR-type
models which originally motivated these experiments.

\

\centerline{\bf Acknowledgments}

\bigskip

We would like to thank Kai Zuber for provoking a re-examination of multi-scalar
modes in double beta decay. One of us (P.B.) wishes to thank the Physics
Department, McGill University
for its warm hospitality while part of this work was being carried out.
We would like to acknowledge research support from N.S.E.R.C.\ of
Canada, les Fonds F.C.A.R.\ du Qu\'ebec, the U.S. National Science Foundation
(grant PHY-9119745) and the Swiss National Foundation.

\listrefs

\figurecaptions

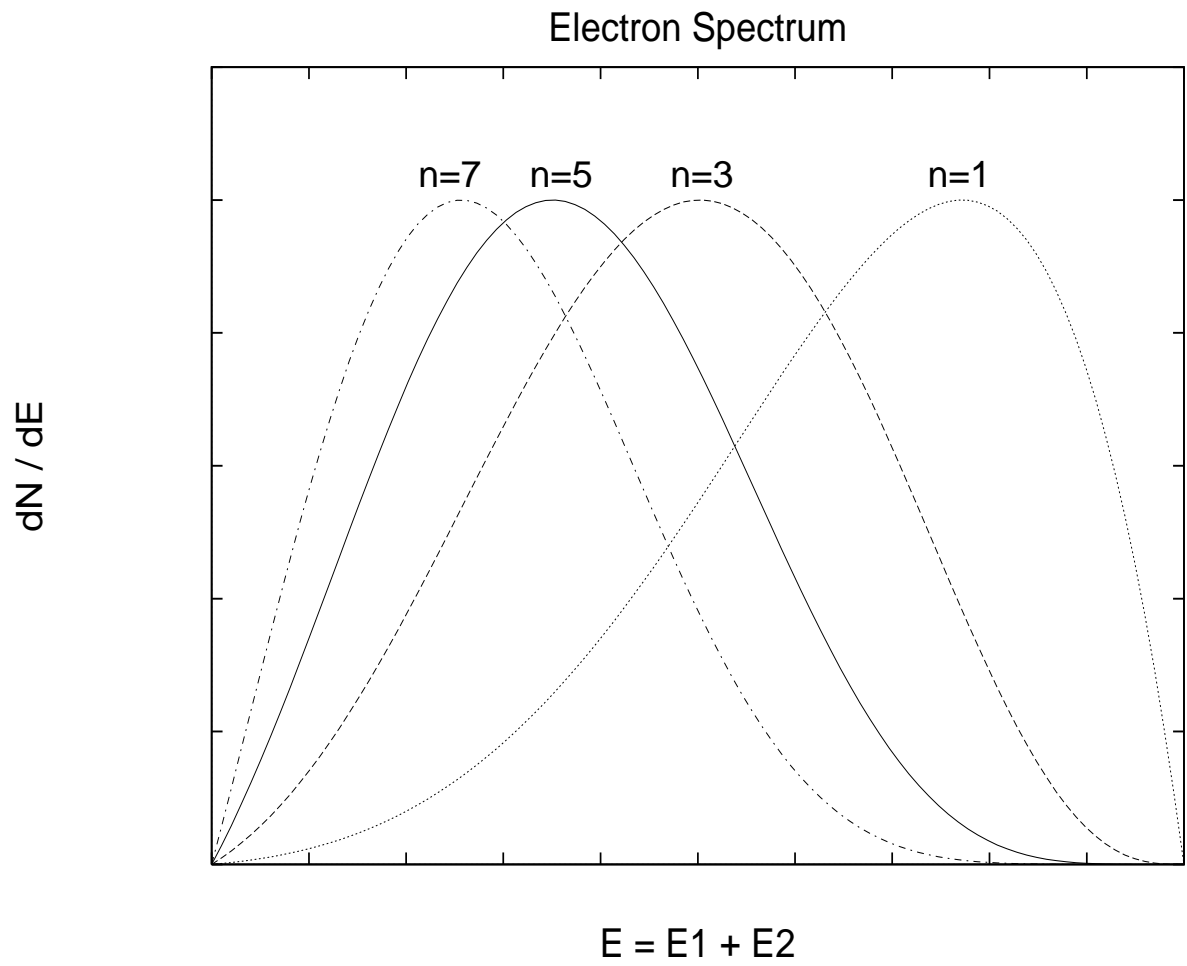
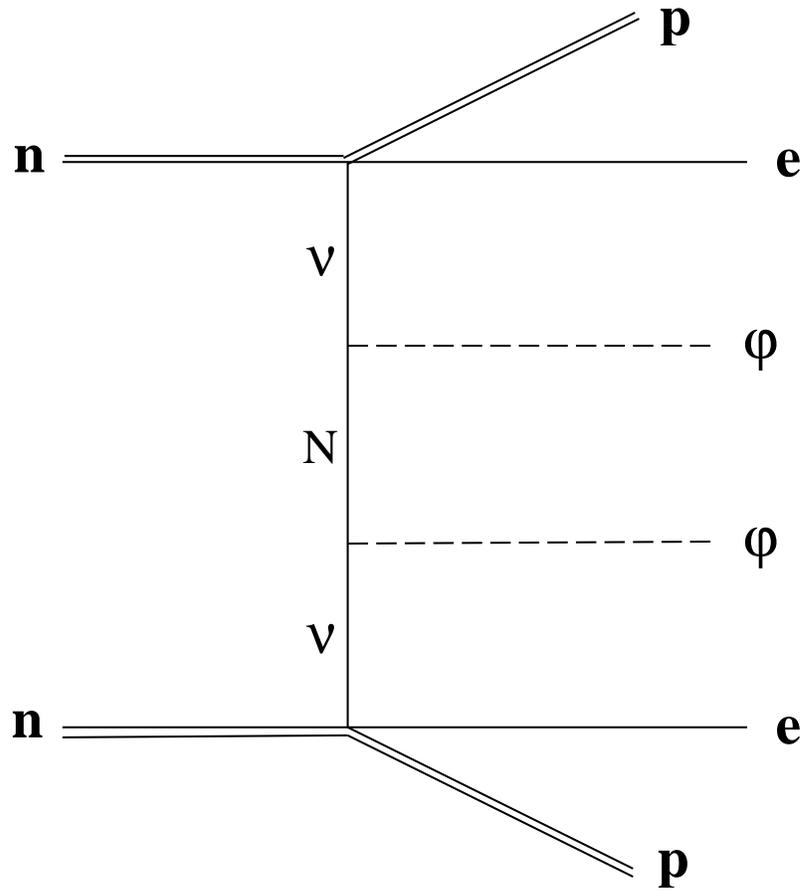
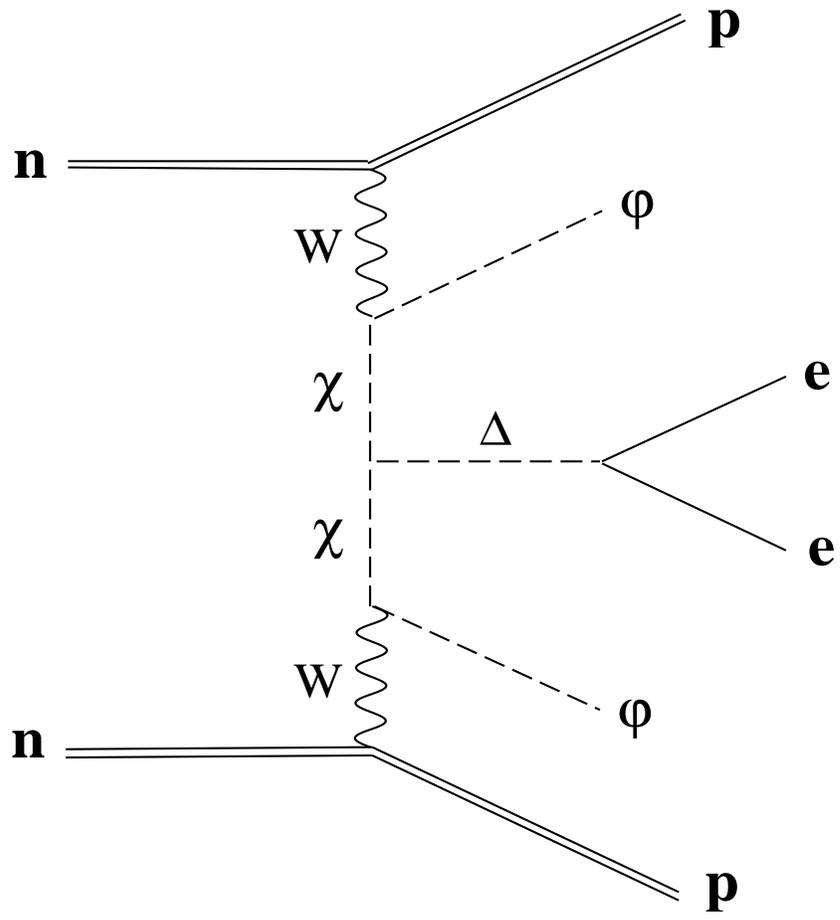
\bye